\documentclass[reprint,amsmath,amssymb,aps,superscriptaddress,prb]{revtex4-2} 

\usepackage{amsmath}
\usepackage{amssymb}
\usepackage{amsthm}
\usepackage[cjk]{kotex}
\usepackage{graphicx}
\usepackage{xcolor}
\usepackage{comment}
\usepackage{ulem}
\usepackage{footmisc}
\usepackage{microtype}
\usepackage{hyperref}
\usepackage{mathtools}
\usepackage{booktabs}

\renewcommand{\v}{\textbf}

\newtheoremstyle{non-italic} 
  {}
  {} 
  {\normalfont} 
  {} 
  {\bfseries}
  {.} 
  {.5em} 
  {}

\begin{document}

%%TC:ignore
\title{Entropy-driven transitions between extended integer and \\ fractional quantum  Hall regimes}

\author{Kyung-Su Kim (김경수)}
\altaffiliation{ \href{mailto:kyungsu@illinois.edu}{kyungsu@illinois.edu}
}
\affiliation{Department of Physics and Anthony J. Leggett Institute for Condensed Matter Theory, University of Illinois  Urbana-Champaign, 1110 West Green Street, Urbana, Illinois 61801, USA} 
\author{Steven A. Kivelson}
\affiliation{Department of Physics, Stanford University, Stanford, CA 94305, USA}

\date{\today}

\begin{abstract}
Electronic states with coexisting Wigner-crystal order can  sometimes exhibit  a quantum  Hall effect over a finite range of electron densities---i.e., exhibit ``extended'' quantum Hall (QH) plateaus in the absence of disorder. Such an extended quantum Hall state can then compete with other QH states over the same density range, allowing a first-order thermal transition between them. Here, we analyze several settings in which the entropy associated with Goldstone modes (e.g., magnons or phonons) or soft gapped modes (e.g., magnetoroton) drives a finite-temperature transition between competing QH regimes.  Applying this framework to moir\'e rhombohedral graphene, we argue that a  soft magnetoroton in a fractional quantum anomalous Hall state provides a plausible bulk mechanism for the observed thermal evolution from an extended integer QH to a fractional  QH regime.
\end{abstract}
\maketitle

%TC:endignore

Among the most exciting recent developments in two-dimensional (2D) materials is the discovery of a plethora of quantum Hall (QH) phases, including zero-field quantum anomalous Hall (QAH) states~\cite{serlin2020intrinsic,li2021quantum,cai2023signatures,zeng2023thermodynamic,park2023observation,xu2023observation,lu2024fractional}. More recently, effectively compressible QH states have been observed~\cite{lu2025extended,butler2026fractional},  in which the quantized Hall response  persists over a  range of densities, seemingly without relying on disorder; these are referred to as  ``extended'' QH states. (Other examples of extended QH states  include  quasiparticle Wigner crystals~\cite{kim2021quantum}, the field-induced spin-density wave phases  seen in certain bulk  molecular crystals~\cite{hannahs1989quantum}, anomalous Hall crystals~\cite{zhou2024fractional,dong2024anomalous}, and ``defect quantum Hall'' states in which mobile interstitial or vacancy defects of a self-doped Wigner crystal~\cite{kim2024dynamical} form a QH fluid~\cite{han2026evidence}.) This allows situations in which one extended QH state  competes with another QH state at the same density.

Specifically, recent experiments in moir\'e rhombohedral graphene (mRG) revealed a thermal competition between  extended integer QAH and fractional  QAH regimes: upon heating, the extended integer QAH state gives way to fractional QAH regimes at several Jain fillings~\cite{lu2025extended}. Existing theory proposals  attribute this thermal evolution to (1) disorder-assisted mechanisms~\cite{das2024thermal,huang2026impurity} or (2) entropy associated with edge excitations~\cite{shavit2024entropy,wei2025edge}.  Here, we develop alternative bulk-entropy-driven mechanisms. (The possibility of entropic stabilization of the fractional QAH state at higher temperatures was also discussed in Ref.~\cite{patri2024extended}.) We analyze several settings in which entropy from gapless Goldstone modes (magnons and magnetophonons) and/or soft gapped modes (magnetorotons) drives a first-order thermal transition between competing quantum Hall states. We then assess their relevance to mRG and argue that, among those considered here, the magnetoroton mechanism is the most plausible for the observed transition. We first outline the general framework and then turn to specific cases.

\smallskip
{\bf General Considerations}---Consider two competing QH states, QH1 and QH2, separated by a zero-temperature first-order transition at $g_c(T= 0)$, with QH1 (QH2) favored for $g< g_c(0)$ ($g>g_c (0)$).  Here,  $g$ may represent a displacement field, an interaction strength, etc. Near the transition, the energy-density difference can be expanded as
$\mathcal E_1(g) 
- 
\mathcal E_2 (g) 
\approx 
\Lambda 
[g-g_c(0)]$,
where $\Lambda >0$. We  now ask how the first-order phase boundary $g_c(T)$ shifts  with increasing temperature.

The relevant quantity at $T>0$ is the free-energy density
$f_a(g,T) 
= 
\mathcal E _a(g) 
+ 
\delta f_a(T)$,
where $\delta f_a (T)$ is the thermal contribution from the low-energy excitations of state $a=1,2$. To leading order, the finite-temperature phase boundary is given by
\begin{align}
\label{eq:g_c}
    g_c(T) 
    - 
    g_c(0) 
    =  
    \frac
    {\delta f_2 (T) - \delta f_1 (T)  }
    {\Lambda} + \ldots,
\end{align}
thermally favoring QH2 over QH1 when $\delta f_1(T) >\delta f_2 (T)$. Consequently, for $g= g_0 < g_c(0)$ sufficiently close to $g_c (0)$, a first-order thermal transition  from the QH1 ground state to  QH2  occurs at  $T_c(g_0)$ determined by
$\delta f_2 (T_c) 
- 
\delta f_1 (T_c)
=
\Lambda [g_0 -g_c(0)]$.
Note that, since topological order does not survive at any finite temperature in two dimensions (2D), this transition should not be interpreted as  a topological phase transition. Rather,  the transition is between the finite-temperature descendants of the two QH states, with different Hall conductances, which may or may not share the same symmetry/ordering.

An important  feature arises when the dominant low-energy excitations are gapless Goldstone modes (e.g., magnons or phonons) associated with the spontaneous breaking of a continuous symmetry in the ground state.  For a Goldstone mode with low-energy dispersion $\omega_a (\v k ) = A_a k^{p_a }$, the leading low-$T$ free-energy density takes the form $\delta f_a (T)  \approx   -  \gamma_a T^{1+ 2/ p_a }$, where $\gamma_a > 0$ depends on the stiffness $A_a$ (see Appendix \ref{app:free-energy}). Thus, if the higher-energy state has a softer dispersion---either a larger exponent $p_a$, or for equal $p_a$, a smaller stiffness $A_a$---it has an entropic advantage and can become  thermodynamically favored upon heating. A gapped collective mode can play a similar role when its gap is comparable to or smaller than the relevant temperature scale.

\begin{figure}
    \centering
    \hspace*{-0.03\linewidth}
    \includegraphics[width=0.9\linewidth]
   {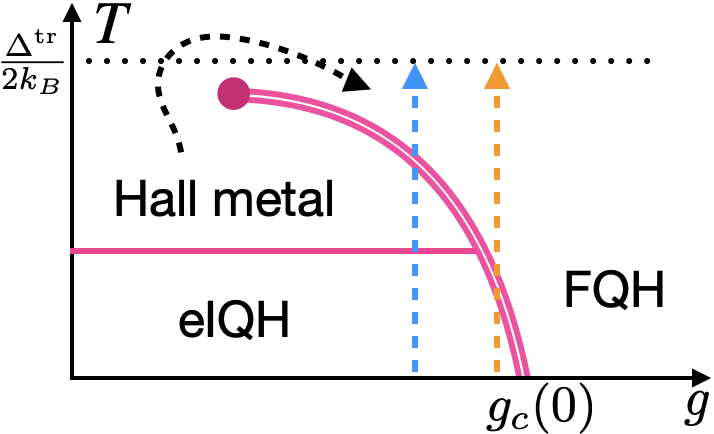}
    \hspace*{0.03\linewidth}
    \caption{
    Schematic $T$-$g$ phase diagram for competing eIQH and FQH regimes. The first-order phase boundary  $g_c(T)$ (double magenta line) bends so as to expand the range of the higher entropy QH regime.  In the case shown, this is the FQH regime, as in cases (i)--(iii) and (vi) in the main text, but it would bend the opposite way in cases (iv)--(v-a). Above the melting boundary (single magenta line), where the translational order is lost, the approximate integer quantization of the Hall response is generically lost, giving rise to a Hall-metal regime. The eIQH-to-FQH evolution can thus proceed either directly (orange dashed line) or through an intermediate Hall-metal regime (blue dashed line). The finite-$T$ Hall response is nearly quantized in the eIQH and FQH regimes, provided  $\Delta^{\rm tr}_{I,F}/2k_B\gg T$, where $\Delta^{\rm tr}_{I,F}$ are the transport gaps of the eIQH and FQH states, respectively. If there is no further  order that differentiates the Hall-metal and the FQH regimes, they can be adiabatically connected without crossing a phase transition, as illustrated by the black dashed line. 
    }
    \label{fig:phase-diagram}
\end{figure}

In this work, we will treat situations in which one of the two competing QH states is ``extended,'' while the other is a translationally-invariant incompressible fluid.  Importantly, while the incompressible fluid can exist  only at a single special rational filling, $\nu=\nu^\star$, of a Landau level (LL) or a Chern band (CB), the extended QH state can persist over multiple values of $\nu$ while retaining the same quantized Hall conductance. In the continuum, this happens when a QH state has a coexisting crystalline order which can accommodate a continuous change in density by adjusting its crystalline periodicity;  such a state is compressible. Importantly, since the crystalline order can be pinned by the edges of the sample even in the absence of disorder, this ``compressible'' QH state supports the quantized Hall response over an extended range of $\nu$~\cite{kim2021quantum}. However, in the presence of a lattice potential, as in the Chern-band case, additional commensurability effects arise. At a discrete set of densities, the electron crystal will commensurately lock to the underlying lattice at sufficiently low temperatures, rendering the state incompressible.   This discrete set of commensurately locked incompressible QH states still constitutes an extended QH regime. (There are several possible scenarios for how the system evolves away from these discrete values of $\nu$---an issue we leave for future work.) We will use ``extended QH regime'' to designate both a compressible QH state that persists over a continuous range of densities and a family of closely related incompressible QH states occurring at a discrete set of densities.

In practice, the commensurability-induced pinning gaps associated with sufficiently high-order commensurabilities are exponentially small, and are thus irrelevant above a correspondingly small  temperature scale. Weak disorder can also wash out such high-order commensurability effects, replacing the discrete set of incompressible QH states by an effectively  compressible QH state extending  continuously over a finite density range. Furthermore, weak disorder rounds the clean first-order boundary in Fig.~\ref{fig:phase-diagram} into  a crossover line when the two finite-$T$ phases are not distinguished by symmetry or ordering.

We now illustrate the mechanisms of entropy-driven QH transitions in Landau levels (LL) or in Chern bands (CB). 
% The various cases are summarized in Table~\ref{table:mode-summary}.

\smallskip
{\bf Spontaneously broken $SU(2)$ symmetry}---Consider a system at a rational filling $\nu^* =  p/q<1$ of a spin-degenerate LL or CB  where an incompressible, spin-polarized fractional QH (FQH)  state~\cite{laughlin1983anomalous,jain1989composite} with Hall conductance $\sigma_H=(e^2/h)(p/q)$  competes with an extended integer QH  state (eIQH) with $\sigma_H = e^2 /h$. We neglect Zeeman splitting and spin-orbit coupling and assume exact $SU(2)$ spin-rotation symmetry. We restrict attention to fillings and parameter regimes in which the FQH state is spin polarized~\footnote{
Within a single LL, the $\nu = 1 $ integer QH ferromagnet is an exact ground state of any spin-independent density-density interaction  $V(\v q ) : \bar \rho (\v q ) \bar \rho (-\v q ): $ with $V(\v q) \geq 0$, where $::$ denotes  normal ordering and $\bar \rho(\v q )$ is the single-LL projected density operator~\cite{bychkov1981two,kallin1984excitations,sondhi1993skyrmions}. At fractional fillings,  interactions can likewise stabilize spin-polarized FQH states, although competing partially polarized or spin-singlet states may occur depending on the filling and microscopic parameters~\cite{park1998phase,davenport2012spinful}. In a Chern band, ferromagnetism is less universal but occurs  over broad parameter  ranges~\cite{zhang2019twisted,bultinck2020mechanism,repellin2020ferromagnetism}.
}.
The resulting FQH ferromagnet breaks $SU(2)$ to $U(1)$ and supports a quadratic magnon with dispersion $\omega = J k^2$~\cite{watanabe2012unified,hidaka2013counting,watanabe2020counting}. For Coulomb interactions in a LL,  $J \sim \ell_B  {e^2 } /{(4\pi\epsilon \hbar)}$  ($\ell_B = \sqrt { {\hbar }/{eB}}$ is the  magnetic length)~\cite{bychkov1981two,kallin1984excitations}.

The eIQH state at the same $\nu^*$, by contrast, arises by doping a $\nu=1$ QH ferromagnet with a concentration $\delta \nu \equiv 1-\nu^*$ of quasiholes.   The lowest-energy charged excitations  in the LL problem (and in a CB as well under appropriate circumstances)  at $\nu=1$ carry skyrmionic spin textures~\cite{sondhi1993skyrmions,fertig1994charged,macdonald1996skyrmions,nayak1996quantum,chatterjee2020symmetry,kwan2022skyrmions,fertig1997hartree}, so for  $\delta \nu$ not too large, it is natural for them to form a skyrmion crystal with zero net moment~\footnote{
Skyrmion, meron, and biskyrmion crystals have been found in LL studies 
\cite{brey1995skyrme,cote1997collective,lilliehook1997quantum,sankararaman2003quantum,cote2007collective}.  Analogous spin/pseudospin skyrmion crystals in Chern bands are also reported~\cite{bomerich2020skyrmion,kwan2022skyrmions}.}.  Such a ``spin-textured integer QH crystal'' supports a non-coplanar spin texture  that completely breaks the $SU(2)$ spin symmetry~\cite{sankararaman2003quantum}.  The three broken generators then produce three linear antiferromagnetic magnons, $\omega_i  = v_i k $   $(i= 1,2,3)$~\footnote{If the state instead carries a nonzero net spin polarization, there will be only two Goldstone modes with one linear $\omega \sim k $ and one quadratic $\omega \sim k^2 $ dispersion \cite{watanabe2012unified,hidaka2013counting,watanabe2020counting}.
}.

The low-temperature free-energy densities of the integer (I) and fractional (F) states are thus
\begin{align}
\label{eq:free-energy_IQH}
    f_I (T) &= \mathcal E_I + f_{I}^{\rm ph}(T)  + f_{I}^{\rm sw}(T)  + O(e^{-{\Delta_I ^{\rm th}}/{k_BT}}),\\
    f_F (T) &= \mathcal E_F  + f_{F}^{\rm sw}(T) + O(e^{- {\Delta_F^{\rm th}}/{k_BT}}),
    \label{eq:free-energy_FQH}
\end{align}
respectively, where $\mathcal E_a $ is the zero-temperature  energy density of state $a= I,F$,   $f^{\rm sw}_{a}$ is the spin-wave contribution to the free energy and  $f^{\rm ph}_{I}$ is the magnetophonon contribution from the crystalline order. The last terms are  activated contributions from other gapped modes controlled by the smallest thermodynamic gap $\Delta_a ^{\rm th}.$

The spin-wave contributions to the free energy are 
\begin{align}
\label{eq:spin-wave-free-energy-integer_spin-degenerate}
    f^{\rm sw}_I &
    \approx
    -C_1 {(k_BT)^3} 
    \sum_{i=1}^3 \frac 1 {(\hbar v_i)^2 } 
    = 
     -C_1  
   \frac {(k_BT)^3} {(\hbar v)^2 } 
    ,
    \\
\label{eq:spin-wave-free-energy-fractional_spin-degenerate}
    f^{\rm sw}_F 
    &\approx
    -C_2 \frac{(k_BT)^2}{\hbar J }.
\end{align}
Here, $C_1 \approx  0.1913$,  $C_2 = \pi/24  $ and $1/v^2 = \sum_{i=1}^3  1/v_i^2 $; the general coefficient $C_p =\Gamma(1 + \frac 2 p)\zeta(1 + \frac 2 p)/(4\pi) $ is derived in  Appendix \ref{app:free-energy}.  The magnetophonon contribution $f_{\rm ph} ^I$ similarly follows from its asymptotic low-energy dispersion $\omega^{\rm ph}_{p}  \sim k^ p$, which in turn depends on the nature of interactions and whether the system arises from  a continuum LL or a CB (see Appendix \ref{app:magneto-phonon-plasmon}):
\begin{align}
\label{eq:magnetophonon-dispersion}
    \omega^{\rm ph}_{p}
    (\v k) 
    \approx
    \begin{cases}
    A_{\frac 3 2} 
    \, k^{\frac 3 2 } 
    & 
    \text{(Coulomb \& LL; $p\equiv \frac 3 2 $)}
    \\    
     A_{2} \, 
    k^{2} 
    & 
    \text{(short-range \& LL; $p \equiv 2$)}
    \\
    \omega _{\rm pin }
    + \ldots 
    &
    \text{(commensurate)}
    \end{cases},
\end{align}
where $A$'s are the stiffness coefficients. In the presence of an underlying crystalline lattice, a (possibly small) gap, $\hbar \omega_{\rm pin}$, is opened in the magneto-phonon spectrum due to commensurability pinning. The corresponding magnetophonon free-energy density is
\begin{align}
\label{eq:magnetophonon-free-energy}
    f^{\rm ph }_I 
    \approx\begin{cases}
           - C_{3/2}
           \frac
           {(k_BT)^{ 7/ 3}}
           {(\hbar A_{3/2})^{4/ 3} }  
           & \text{(Coulomb \& LL)} 
           \\
          - C_2  
           \frac{(k_B T)^{2}}{\hbar A_{2} } 
           & \text{(short-range \& LL)}
           \\
           O(e^{-\hbar \omega_{\rm pin}/k_BT })
           &
           \text{(commensurate)}
    \end{cases},
\end{align}
where $C_{3/2 } \approx 0.1341$. We discuss these three cases separately in  (i)--(iii) below.

\smallskip
(i) {\it Coulomb interaction in a spin-degenerate LL.} 
For an unscreened Coulomb interaction, the $O(T^2)$ quadratic magnon contribution of the ferromagnetic FQH state dominates over the $O(T^{7/3},T^3)$ contributions of the   spin-textured eIQH state at  low $T$,  so
\begin{align}
\label{eq:Delta_f_Coulomb}
    f_F-f_I =
    \mathcal E_F - \mathcal E_I 
    - C_2 
    \frac{(k_B T)^{2}}{\hbar J } 
    + \ldots.
\end{align} 
i.e., the ferromagnetic FQH state is entropically favored. Thus, if the eIQH is the ground state, $\mathcal E_F >\mathcal E_I$, and the two states are sufficiently close in energy, the transition to the FQH regime occurs at 
\begin{align}
\label{eq:Tc_Coulomb_spin-degenerate}
    k_BT _c \approx 
    \sqrt{\frac {24\hbar J }{\pi}
    }{\ \sqrt{ \mathcal E_F - \mathcal E_I }}\ .
\end{align}
The validity of this expression is discussed in Appendix~\ref{app:validity}.

%\ksk{
The first-order transition at $T_c$ may involve a change in crystalline order. If, at a given parameter, $T_c$ is smaller than the melting temperature $T_{\rm melt}$ of the eIQH crystal, the first-order transition preempts melting and destroys quasi-long-range translational order discontinuously. If $T_c > T_{\rm melt}$, by contrast, the eIQH crystal has already melted before reaching the first-order boundary. Consequently, two scenarios are possible: 
(a) The quasi-long-range translational order of the eIQH has melted before the first-order line reaches the ordinary critical point (magenta dot in Fig.~\ref{fig:phase-diagram}). Since the crystal  has a finite shear rigidity, it is generically pinned so  the eIQH region below the melting line still exhibits  a nearly integer-quantized Hall conductivity with activated corrections. However, once the quasi-long-range translational order  melts, there is no reason to expect any approximate integer quantization, yielding an unquantized Hall response: we refer to this phase as a ``Hall metal''~\footnote{
In the clean thermodynamic limit, the melted crystal generically contributes nontrivially to the Hall conductivity, spoiling the approximate integer quantization. In the presence of weak disorder, the situation is more subtle: since the quasi-long-range translational order does not survive in the presence of disorder, the melting line is replaced by a crossover between the low-$T$ pinned ``disordered crystal''---which has a large, albeit finite, crystalline correlation length---and a high-$T$ Hall metal. Within the disordered crystalline regime, the Hall response can remain approximately quantized when the disorder-pinning length is shorter than the crystalline correlation length.
}. 
This scenario is illustrated in Fig.~\ref{fig:phase-diagram}.
%(Within the Kosterlitz-Thouless-Halperin-Nelson-Young scenario, the complete melting occurs in two steps, with an intermediate hexatic Hall-metal phase appearing between the crystalline eIQH and the isotropic Hall metal~\cite{halperin1978theory,nelson1979dislocation,young1979melting}. Fig. \ref{fig:phase-diagram} shows the simplest subcase in which the orientational order is also lost before the first-order line reaches the ordinary critical point; if the hexatic order instead persists to the would-be critical point, the first-order line must first meet the hexatic-to-isotropic transition boundary, producing additional multicritical structure.)
(Under some circumstances, there can also be an intermediate hexatic Hall-metal phase, not shown in Fig.~\ref{fig:phase-diagram}.~\cite{halperin1978theory,nelson1979dislocation,young1979melting})
(b) On the other hand, the quasi-long-range translational order of the eIQH state could persist toward the would-be ordinary critical point. In this case, the first-order line cannot simply terminate there; instead, it must meet the melting boundary and produce additional multicritical structures.

There will generically be a discontinuity in the Hall resistance across the first order line, but its magnitude is scenario-dependent.  Along the orange dashed line in Fig. \ref{fig:phase-diagram}, the Hall conductivity  would likely jump between nearly quantized integer and fractional values  so long as $k_BT \ll \Delta^{\rm tr}_{I,F}/2 $, where $\Delta_a^{\rm tr} =  \epsilon_{\rm qp}^{(a)} + \epsilon_{\rm qh}^{(a)}$ is the transport gap.  In contrast, if the thermal evolution from the eIQH to the FQH passes through an  intermediate Hall-metal regime, as illustrated by the dashed blue line in Fig. \ref{fig:phase-diagram}, the jump in the Hall conductivity will tend to be much smaller. In either case, when the FQH state is spin polarized, the transition should also  result in  a large jump in the uniform ferromagnetic susceptibility, reflecting the exponentially large spin-correlation length in the FQH regime, $\xi_F^{\rm sp} \sim \xi_F^{(0)}  e^{\rho_{F}/k_BT }$, where $\rho_F$ is the appropriate (energy-valued) spin stiffness  and $\xi_F^{(0)}$ is a microscopic length scale~\cite{read1995continuum}.

\smallskip
(ii) {\it Short-range interactions in a spin-degenerate LL.} 
When the Coulomb interaction is screened by a nearby metallic gate, the long-wavelength magnetophonon dispersion of the eIQH is quadratic, $\omega  = A_2  k^2$~\eqref{eq:magnetophonon-dispersion}. Its contribution to the free energy therefore has the same  $T^2$ scaling as the ferromagnetic-magnon contribution, so 
\begin{align}
\label{eq:Delta_f_short-range_LL}
   \!\!f_F \!- f_I \!= 
   \mathcal E_F - \mathcal E_I 
           & + 
           \frac \pi {24} 
           {(k_B T)^{2}} 
           \!\left (
           \frac 1 {\hbar A_2 }- 
           \frac 1 {\hbar J }
           \right )
           + \ldots.
\end{align}
Unlike case (i), there is no universal entropic preference: the FQH (eIQH) regime is favored when $J<A_2$ ($J>A_2$).

A thermal transition occurs when the $T=0$ ground state is the one disfavored entropically.  If $\mathcal E_F > \mathcal E_I  $ and $J<A_2$, there is a thermal transition from the low-$T$ eIQH regime to the high-$T$ FQH regime at 
\begin{align}
\label{eq:critical-temperature_short-range}
    k_B T_c 
    \approx 
    \sqrt{
    \frac{24 \hbar J  }{\pi }  
    \left(\frac{A_2}{A_2 - J }\right)
    \left ( \mathcal E_F - \mathcal E_I \right ) }
    .
\end{align}
Conversely, if $\mathcal E_F < \mathcal E_I  $ and $J > A_2$, heating drives the reverse transition from the FQH to the eIQH regime at $T_c$~\eqref{eq:critical-temperature_short-range}, subject to the validity conditions discussed in Appendix~\ref{app:validity}. In the latter case, if it happens that $T_c $ is smaller than the melting temperature $T_{\rm melt}$ of the eIQH crystal,  this implies that quasi-long-range translational order arises upon heating above $T_c$.

\smallskip
(iii) {\it Spin-degenerate Chern band.}
Because the crystalline order in the eIQH state is assumed to be commensurately pinned by the underlying lattice, its  phonon contribution is activated~\eqref{eq:magnetophonon-free-energy}, and the  free-energy difference is therefore 
\begin{align}
\label{eq:Delta_f_short-range_ChernBand}
   f_F - f_I = 
   \mathcal E_F - \mathcal E_I 
           & - 
           \frac \pi {24} 
           \frac {(k_B T)^{2}} {\hbar J }
           + \ldots .
           % O(T^3, e^{-\hbar\omega_{\rm pin}/k_BT}).
\end{align}
Then, as in case (i), the leading $O(T^2)$ contribution favors the fractional Chern insulator. If the eIQH is the ground state, heating can thus drive a transition to the FQH with  analogous expression for $T_c$ as in \eqref{eq:Tc_Coulomb_spin-degenerate}, subject to the validity conditions discussed in Appendix~\ref{app:validity}. The phonon is thermally suppressed when $k_BT_c \ll \hbar \omega_{\rm pin}$. For weaker pinning, its contribution must be retained~\cite{footnote:phonon-Chern-band}, which modifies $T_c$ and, for short-range interactions, can reverse the entropic preference.

Because the crystalline order of the commensurate eIQH breaks only a discrete lattice-translation symmetry, its long-range crystalline order can survive at finite  $T_{\rm melt}$. Hence, if $T_c<T_{\rm melt}$, the first-order eIQH-to-FQH transition simultaneously restores  lattice-translation symmetry, which is the case illustrated by the orange dashed line in Fig.~\ref{fig:phase-diagram}. On the other hand, if $T_c>T_{\rm melt}$, the thermal evolution proceeds in two steps, eIQH$\to$Hall metal$\to$FQH, as illustrated by the blue dashed line.

\smallskip
(iv)--(v) {\it Spin-split Landau level/Chern band}---When a Zeeman field or Ising-like spin-orbit coupling reduces the spin symmetry from $SU(2)$ to $U(1)$, the preceding conclusions are modified. The spin-wave excitations of the spin-polarized FQH are now gapped. In a LL [case (iv)], eIQH retains a gapless magnetophonon, whose free-energy contribution favors the integer regime. In a  CB [case (v)], the phonon is gapped when the crystal is commensurately pinned.  Assuming the eIQH state still involves  a non-coplanar spin texture [case (v-a)], it retains  one gapless linear  magnon mode associated with spontaneous $U(1)$ spin-symmetry breaking, whose $O(T^3)$ entropy favors the integer regime. By contrast, for a fully spin-polarized eIQH state [case (v-b)], both competing states are gapped and there is no universal low-$T$ entropic preference.  More detailed discussions of these cases can be found in Appendix~\ref{app:spin-split}.

\smallskip
{\bf Soft gapped collective modes}---Gapped collective excitations can also give an appreciable thermodynamic contribution when their gap is comparable to or smaller than $k_BT$. A particularly favorable example is the FQH magnetoroton mode~\cite{girvin1986magneto}, whose finite-momentum minima form a ring in a rotationally invariant LL, enhancing the low-energy phase space and hence the entropy. In a CB, lattice anisotropy generally warps this ring into a finite set of symmetry-related minima~\cite{shen2026magnetorotons,kousa2025theory,long2026spectra}; nevertheless, if the associated warping energy $W \ll k_BT$, the ring-minimum description remains a good approximation. We now analyze this scenario in detail.

\smallskip
(vi) {\it Magnetoroton.} Magnetoroton entropy was previously shown to drive finite-$T$ competition between Laughlin liquids and Wigner crystals~\cite{platzman1993quantum,price1993freezing}. Here, we extend this mechanism to describe the competition between an eIQH and a translation-invariant FQH state at a rational filling $\nu=\nu^\star$. We  assume that the magnetoroton is the only thermally accessible mode in the FQH state: spin excitations are gapped,  either intrinsically  in a gapped spin-singlet state~\cite{halperin1983theory,ardonne1999new}, or by Zeeman coupling or SOC in a polarized state. We first  take the crystalline order of the eIQH state to be pinned, and later relax this assumption to include the phonon contribution. We approximate the magnetoroton dispersion near its minimum as
\begin{align}
    \epsilon_R (k ) 
    =
    \Delta _R  
    +
    \frac 
    {\hbar^2 (k-k_R)^2}
    {2m_R } 
    +\ldots 
\end{align}
where $\Delta_R $ is the gap, $k_R$ is the radius of the ring, and $m_R$ is the radial effective mass. For $W\ll k_BT$, we neglect  the magnetoroton warping effects, and further require 
$k_B T 
\ll 
E_R 
\equiv 
{\hbar^2 k_R^2}/{(2m_R)}$
so that thermal fluctuations probe only a narrow radial region around $k= k_R$. 

Under these conditions,  the magnetoroton free-energy density, within the harmonic approximation, is~\footnote{
Here, we treat magnetorotons as noninteracting bosonic quasiparticles, which is justified in the dilute regime $k_BT \ll \Delta _R $. When $k_BT \sim \Delta_R$, magnetoroton interactions can give non-negligible corrections to $f_R$, which we neglect here.
}
\begin{align}
    &f_R 
    \approx 
    - \gamma_R (k_BT)^{3/2}
    {\rm Li}_{\frac 3 2} 
    (e^{-\frac{\Delta_R}{k_B T}}),
    &
    \gamma_R \equiv \frac{k_R \sqrt {m_R}}{\hbar\sqrt {2\pi }}.
\end{align}
Here,  ${\rm Li}_{s}(z) \equiv \sum_{n=1}^\infty  z^n / n^s $ is the polylogarithm (see Appendix \ref{app:magnetoroton} for details). The $T^{3/2}$ prefactor reflects the enhanced low-energy density of states associated with the ring of minima. In the low-temperature limit $k_BT \ll \Delta_R $, this reduces to 
$f_R 
\sim 
- \gamma_R (k_BT)^{3/2}
e^{-\Delta_R /k_B T}$. Of particular interest is the case in which the integer state is the ground state. Upon heating, the transition to the fractional regime occurs at $T_c$ determined by
\begin{align}
\label{eq:Tc-magnetoroton}
    \mathcal E_F -\mathcal E_I
    = 
    \frac{k_R \sqrt {m_R}}{\hbar\sqrt {2\pi } }
    (k_B T_c)^{3/2 }
    {\rm Li}_{3/2 }
    (e^{-\Delta _R /k_BT_c }).
\end{align}
At fixed $k_R$ and $m_R$, $T_c$ is lowered as the magnetoroton gap $\Delta_R$ softens or as the  energy-density difference $\mathcal E_F - \mathcal E_I $ decreases.

The above discussion neglects the phonon contribution of the integer crystal, which is justified for strong pinning, $\hbar \omega_{\rm pin} \gg k_BT$. If the integer crystal is weakly pinned or unpinned,  the phonon contribution $f^{\rm ph}_I$ must also be retained \cite{footnote:phonon-Chern-band}.
The transition condition  then becomes 
$\mathcal E_F - \mathcal E_I
=f_I^{\rm ph}(T_c) - f_R(T_c)$.
Because $f_I^{\rm ph}(T) - f_R(T)$ can be a nonmonotonic function of $T$,  multiple solutions are possible for either sign of $\mathcal E_F - \mathcal E_I$, allowing   reentrant FQH--IQH--FQH (for $\mathcal E_F < \mathcal E_I$) or IQH--FQH--IQH (for $\mathcal E_F > \mathcal E_I$) sequences upon increasing temperature.

\smallskip
{\bf Application to mRG}---Recent bulk-sensitive measurements~\cite{li2026competing} provide compelling evidence that the observed thermal transition from an anomalous eIQH to FQH at $T_c \sim 100$--$340\,  {\rm mK}$~\cite{lu2025extended} reflects a bulk phase transition.  Among the Goldstone-mode mechanisms, case (iii) has the necessary direction of the transition under the assumption that the FQH state is spin-polarized~\footnote{We caution  that the spin polarization of the FQH states in mRG remains unsettled~\cite{dong2024theory}.}.  In the mRG experiment, however, even if the FQH state is spin-polarized, its magnon is expected to be gapped by spin-orbit coupling (SOC) and  Zeeman coupling to the applied magnetic field. Experimental analyses have extracted an intrinsic SOC scale of  $\lambda _{\rm SOC} \approx 50 \, \mu {\rm eV} \approx 0.58\, {\rm K}$ in rhombohedral trilayer graphene \cite{arp2024intervalley}, while  measurements in tetralayer graphene have been interpreted as implying $\lambda_{\rm SOC} \gtrsim 120\, \mu {\rm eV}$ \cite{auerbach2025isospin}. The $0.1\, {\rm T}$ out-of-plane magnetic field  used in the eIQH experiment \cite{lu2025extended} further adds a Zeeman energy $g \mu_B B \approx 12\, \mu{\rm eV} \approx 0.13\, {\rm K}$ to the intravalley spin gap for the spin-valley polarization favored by the applied field. Using the smaller trilayer value as a  conservative estimate therefore gives an  intravalley spin gap of $\Delta _{\rm sw}/k_B \approx 0.58\, {\rm K} + 0.13\, {\rm K} = 0.71 \, {\rm K}$. Other intervalley collective modes are also expected to be gapped well above the experimental temperature scale; see Appendix~\ref{app:spin-split}. The quadratic-magnon free energy is therefore suppressed from its gapless value by ${\rm Li}_2 (e^{-\Delta_{\rm sw } /k_BT})/\zeta (2)$. For $\Delta_{\rm sw}/k_B \approx 0.71 \, {\rm K}$, this factor is approximately $5 \times 10^{-4}$ at $T= 100\, {\rm mK}$ and $8 \times 10^{-2 }$ at $T= 340\, {\rm mK}$,   making case (iii) less likely to provide sufficient entropy over the experimental temperature range.

\begin{figure}[t]
    \centering
    \includegraphics[width=0.9\linewidth]{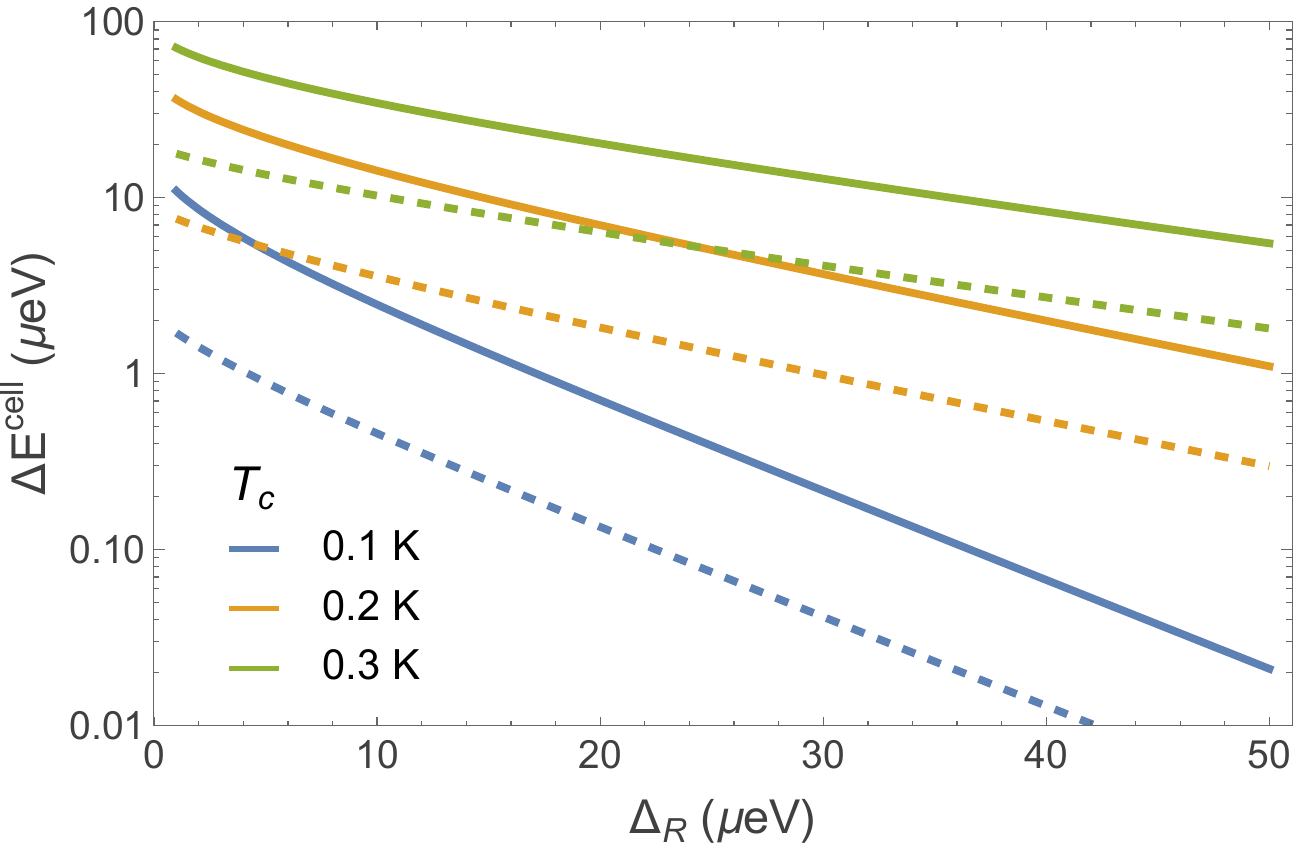}
    \caption{The estimated energy difference $\Delta E ^{\rm cell}$ per moir\'e unit cell between the eIQH and FQH states to give a thermal transition at $T_c$ as a function of the magnetoroton gap $\Delta_R$. Solid lines use the ring-minima approximation~\eqref{eq:magnetoroton-mRG}, whereas dashed lines use the discrete-minima result, Eq.~\eqref{eq-app:magnetoroton-free-energy-discrete}, with $N_R=3$ and $E_\theta\equiv\hbar^2k_R^2/(2m_\theta)=E_R$.
    }
    \label{fig:magnetoroton}
\end{figure}

This motivates us to consider the soft-magnetoroton mechanism. To test the plausibility of this scenario, we make crude estimates of the various parameters that enter the analysis as follows:  Using \eqref{eq:Tc-magnetoroton}, we give an order-of-magnitude estimate of the energy difference between the integer and fractional states required to produce a transition at $T_c \sim 100$--$340{\rm mK}$ \cite{lu2025extended,li2026competing}. We set $k_R$ to the moir\'e Brillouin-zone scale by defining $\pi k_R^2  = (2\pi)^2/A_M$, where $A_M= \frac{\sqrt 3 }{2} a_M^2 $ is the area of the moir\'e unit cell and $a_M \approx 11.5\, {\rm nm}$ is the moir\'e lattice constant~\cite{lu2024fractional}. We choose  $m_R$ such that $E_R = \hbar^2 k_R ^2/2m_R$ is of order the transport gap $\Delta^{\rm tr}$---a characteristic excitation scale. At $\nu=2/3$, compressibility measurements find a chemical-potential jump---an energy cost to add one electron---of $\Delta\mu\approx0.5\,{\rm meV}$~\cite{butler2026fractional}. In the clean dilute-quasiparticle limit, this corresponds to $\Delta ^{\rm tr}\sim (e^*/e)\Delta \mu  \approx  0.17 \, {\rm meV} $, where $e^* = e/3$ is the quasiparticle charge. Thus, with an estimated $E_R \sim  0.17 \, {\rm meV} $,  the required  energy difference per moir\'e unit cell, $\Delta E^{\rm cell}=(\mathcal E_F-\mathcal E_I)A_M$, is~\footnote{
This estimate neglects eIQH phonon entropy and treats the magnetorotons as noninteracting bosonic quasiparticles with an approximate ring of minima. These require $\hbar \omega_{\rm pin} \gg k_BT_c$ and   $\Delta_R\gg k_BT_c \gg W $, respectively. The strong-pinning assumption is plausible in mRG: recent estimates give an effective moiré potential of order \(9\)--\(12\,{\rm meV}\)~\cite{regnault2026moire}, far above \(k_BT_c\).
The narrow-shell approximation further requires   $k_BT_c \ll E_R = \hbar^2 k_R^2 /(2m_R)$. 
If any of these conditions fail,  \eqref{eq:magnetoroton-mRG} must be modified accordingly; in particular, when $\Delta_R$ is not large compared with \(k_BT_c\), the magnetoroton gas is no longer dilute and the noninteracting approximation is not quantitatively controlled.
}
\begin{align}
\label{eq:magnetoroton-mRG}
    \Delta E^{\rm cell}
    \sim 
    \sqrt{\frac{4\pi}{0.17\,{\rm meV}}}
    (k_BT_c)^{3/2}
    {\rm Li}_{3/2}
    \left(
    e^{-\frac{\Delta_R}{k_BT_c}}
    \right).
\end{align}
To illustrate the effect of lattice warping, we also consider the opposite limit $W\gg k_BT_c$, where the magnetoroton ring splits into discrete minima (see Appendix~\ref{app:magnetoroton} for more detail). For the estimate below, we take $N_R=3$ symmetry-related minima and $E_\theta \equiv \hbar^2 k_R^2 /(2m_\theta) = E_R$.

Fig. \ref{fig:magnetoroton} shows the required energy difference $\Delta E^{\rm cell}$ as a function of $\Delta _R$ in both the ring-minima (solid lines) and discrete-minima (dashed lines) limits. For example, for $\Delta_R=20\,\mu{\rm eV}$ and $T_c = 200 \, {\rm mK} $, the noninteracting estimate within the ring-minima approximation~\eqref{eq:magnetoroton-mRG} gives $\Delta E^{\rm cell}\approx7\,\mu{\rm eV}$.  Although such a small magnetoroton gap---in particular, much smaller than the transport gap $\Delta^{\rm tr}$---is not generic in the conventional LL setting \cite{girvin1986magneto,scarola2000rotons,balram2017positions}, substantial magnetoroton softening can occur near  a crystalline instability~\cite{kousa2025theory,shi2026effects,long2026spectra,lu2026continuous}. Such proximity of the FQH state to a crystalline instability, in addition to its being in close competition with an eIQH state, is a separate microscopic question that can be tested by calculating or measuring the neutral spectrum.

\smallskip
{\bf Implications}---Our analysis suggests several direct  tests of the proposed mechanisms in mRG. First, a realistic calculation of the FQH magnetoroton dispersion in mRG or finite-$\v q$ spectroscopy could determine whether the magnetoroton is indeed sufficiently soft. Moreover, if the FQH state is spin-polarized while the  eIQH state is not, a tilted-field experiment could probe their spin polarization and potentially constrain their energy difference. It is also interesting to investigate whether an analogous mechanism involving low-energy excitations could drive the current-induced eIQH-to-FQH transition observed in the same experiment \cite{lu2025extended}, providing an alternative to the scenario based on electric polarizability \cite{patri2024extended}.

%TC:ignore

\section*{Acknowledgments}
We thank Long Ju, Kin Fai Mak and Luke Kim for helpful comments on the draft. 
ChatGPT models GPT-5.6 Sol and GPT-6 Astra were used for scientific discussion and manuscript editing. The authors carefully reviewed all AI-assisted content and take full responsibility for the manuscript. KSK is supported by the Anthony J. Leggett Postdoctoral Fellowship at the University of Illinois Urbana-Champaign. Part of the work by KSK was performed  at the Aspen Center for Physics, which is supported by a grant from the Simons Foundation (1161654, Troyer). SAK was supported, in part,  by NSF-BSF award DMR2310312 at Stanford.

% \bibliography{main}

% \section{End Matter}

% \iffalse
\iftrue

\appendix
\onecolumngrid

\section{Spin-split Landau level or Chern band}
\label{app:spin-split}

\begin{table*}[t]
\caption{
Summary of Goldstone-mode mechanisms (i)--(v). The last column indicates which state is favored by the leading thermal contribution. $SU(2)$ and $U(1)$ denote the spin symmetries in spin-degenerate and spin-split cases, respectively.  ST-eIQH: spin-textured eIQH. SP-eIQH: fully spin-polarized eIQH.
}
\label{table:mode-summary}
\centering
\small
\renewcommand{\arraystretch}{1.35}
\setlength{\tabcolsep}{6pt}
\begin{tabular}{
@{}
p{0.20\textwidth} %narrowest
p{0.34\textwidth} %narrowest
p{0.22\textwidth} %narrowest
p{0.17\textwidth} %narrowest
@{}
}
\toprule
\textbf{Setting}
&
\textbf{Integer state}
&
\textbf{Fractional state}
&
\textbf{Entropy favors}
\\
\midrule
(i) $SU(2)$ LL, Coulomb
&
3 linear magnons $\omega_i = v_i k$; 
% ($i=1,2,3)$;
\newline
magnetophonon $\omega=A_{3/2}k^{3/2}$
&
Quadratic magnon $\omega=Jk^2$
&
Fractional
\\[2mm]
(ii) $SU(2)$ LL, short-range
&
3 linear magnons $\omega_i = v_i k$;
\newline
magnetophonon $\omega=A_2k^2$
&
Quadratic magnon $\omega=Jk^2$
&
Fractional if $J<A_2$;
\newline
integer if $J>A_2$
\\[2mm]
(iii) $SU(2)$ CB
&
3 linear magnons
&
Quadratic magnon $\omega=Jk^2$
&
Fractional
\\[2mm]

(iv) $U(1)$ LL
&
$\omega=A_{3 /2}k^{3 /2}$ magnetophonon  for Coulomb;
\newline
$\omega=A_2k^2$ magnetophonon for short-range; 
\newline
1 (0) linear magnon for ST-eIQH (SP-eIQH)
&
Gapped
&
Integer
\\[2mm]

(v-a) $U(1)$ CB:
ST-eIQH
&
1 linear magnon $\omega=vk$
&
Gapped
&
Integer
\\ [2mm]

(v-b) 
$U(1)$ CB:
SP-eIQH
&
Gapped
&
Gapped
&
No generic preference
\\

\bottomrule
\end{tabular}
\end{table*}

In cases (i)--(iii) of the main text, we considered the thermal competition between eIQH and FQH states in the presence of exact $SU(2)$ spin-rotation symmetry.  Here, we give the details of the spin-split cases, focusing on situations in which a Zeeman field or an Ising-like spin-orbit coupling (SOC) reduces the spin symmetry to a  residual $U(1)$. We assume that the FQH state is fully spin-polarized, so that its spin excitations are gapped. The eIQH state can take two forms depending on the strength of the spin splitting. (a) For sufficiently weak spin splitting, the lowest-energy charged excitations can remain dressed by non-coplanar spin textures and crystallize, forming a spin-textured eIQH. This state spontaneously breaks the residual $U(1)$ symmetry and supports one gapless antiferromagnetic-magnon mode, $\omega = vk$~\cite{cote1997collective}. For more generic SOC that completely breaks the $SU(2)$ spin-rotation symmetry, this magnon mode is also gapped. On the other hand, (b) for larger spin splitting, the lowest-energy charged excitations become electron- or hole-like rather than spin-textured \cite{sondhi1993skyrmions,fertig1997hartree}, leading to a fully spin-polarized eIQH. This state preserves the $U(1)$ symmetry and has a finite spin gap $\Delta^{\rm spin}_I$.

The leading free-energy difference can then be written as 
\begin{align}
    f_F - f_I 
    =
    \mathcal E_F - \mathcal E_I
    - f_I^{\rm ph}
    - f_I^{\rm sw}
    +\ldots,
\end{align}
where $f_I^{\rm ph }$ and $f_I^{\rm sw}$ are the magnetophonon and spin-wave contributions of the integer state, respectively, and $\ldots $ include other activated corrections. The magnetophonon contribution~\eqref{eq:magnetophonon-free-energy} of the integer state scales as $O(T^2)$ or $O(T^{7/3})$ in a LL and is  activated in the commensurately pinned Chern-band case. The spin-wave contribution $f_I^{\rm sw}$ scales as $O(T^3)$ for the spin-textured eIQH, and is  activated, $O(e^{-\Delta ^{\rm spin }_I/k_BT})$, for the fully spin-polarized eIQH. We now discuss the LL and Chern-band cases separately.

\smallskip
(iv) {\it Spin-split Landau level.}
In a LL, the crystalline order of the eIQH state supports a gapless magnetophonon, giving a power-law free-energy contribution.
\begin{align}
   f_F - f_I = 
     \mathcal E_F
     -\mathcal E_I
           +
           \begin{cases}
            C_{3/2}
           \frac
           {(k_BT)^{ 7/ 3}}
           {(\hbar A_{3/2})^{4/ 3} }  
            &\!\text{(Coulomb)}
           \\
           C_2  
           \frac{(k_B T)^{2}}{\hbar A_{2} } 
           &\!\!\!\!\!\!\!\! \text{(Short-range)}
    \end{cases}
           \!+ \ldots.
\end{align} 
The omitted terms include subleading corrections to the asymptotic magnetophonon dispersion and activated contributions; for a spin-textured eIQH [weak spin-splitting case (a)], they additionally include the $O(T^3)$ antiferromagnetic-magnon contribution. In either case, the magnetophonon entropy favors the crystalline eIQH state. Thus, a nontrivial thermal transition  arises when the FQH state is the ground state, $\mathcal E_I >\mathcal E_F $: heating then drives a transition to the higher-$T$ integer regime at 
\begin{align}
    k_BT_c \approx  
    \begin{cases}
        \frac
        {(\hbar A_{3 / 2})^{4/7}} 
         {C_{3/ 2}^{3/ 7}}
         \left (
        {\mathcal E_I -\mathcal E_F}
        \right )^{\frac 3 7} 
        & \text{(Coulomb)}
        \\
        \sqrt{ 
        \frac{\hbar A_{2}}{C_{2}}
        }
        \sqrt{\mathcal E_I -\mathcal E_F}
        & \!\!\!\!\!\!\! 
        \text{(Short-range)}
    \end{cases},
\end{align}
subject to the validity conditions discussed in Appendix~\ref{app:validity}.

\smallskip
(v) {\it Spin-split Chern band.}
At the commensurate filling of a Chern band, the commensurate locking gaps the phonon mode, giving an activated contribution to the free energy at $k_BT \ll \hbar \omega_{\rm pin}$ \eqref{eq:magnetophonon-free-energy}. The low-temperature competition therefore depends on whether the integer state spontaneously breaks the $U(1)$ spin symmetry.

(v-a) {\it Spin-textured eIQH}.  
If the integer state is spin-textured, it spontaneously breaks the residual $U(1)$ spin symmetry and supports one gapless antiferromagnetic magnon mode, $\omega = v k$. The leading free-energy difference becomes 
\begin{align}
    f_F - f_I = 
    \mathcal E_F
    -\mathcal E_I
    + C_1
    \frac{(k_BT)^{3}}{(\hbar v)^{2}} 
    + \ldots,
\end{align}
and the $O(T^3)$ spin-wave entropy favors the integer regime. If the FQH is the ground state, $\mathcal E_F < \mathcal E_I$, heating drives a transition to the integer regime at 
\begin{align}
\label{eq:Tc-case5}
    k_B T_c 
    = 
    \left [
    \frac {(\hbar v)^2 } {C_1} 
    \right ]^{\frac 1 3 }
    (\mathcal E_I -\mathcal E_F)^{\frac 1 3}.
\end{align}
Note that, unlike the $SU(2)$ case, the spin-textured eIQH at $T=0$ spontaneously breaks only the residual $U(1)$ spin symmetry and can support finite-$T$ quasi-long-range spin order below $T_{\rm BKT}$. Together with the finite-$T$ crystalline order associated with discrete lattice-translation symmetry breaking discussed in case (iii), this further enriches the finite-$T$ phase diagram beyond the one shown in Fig.~\ref{fig:phase-diagram}.

(v-b) {\it Fully spin-polarized eIQH}.
For the fully spin-polarized eIQH, both spin and phonon excitations are  gapped. Since the competing FQH is also assumed to be gapped, 
\begin{align}
   f_F- f_I = 
     \mathcal E_F
     -\mathcal E_I
           + 
           O(e^{-\Delta_{\rm min} /k_BT }),
\end{align}
where $\Delta_{\rm min}$ denotes the smallest relevant excitation gap of the two states. There is therefore no universal power-law entropic preference at asymptotically low temperature; any thermal transition in this regime depends on the relative gaps and degeneracies of the activated modes.

The results in (v-a) and (v-b) assume $k_B T_c \ll \hbar \omega_{\rm pin}$, so that the phonon contribution is thermally suppressed. For weaker pinning or higher temperatures $k_B T \gtrsim \hbar \omega_{\rm pin}$, the appropriate phonon contribution must also be included~\cite{footnote:phonon-Chern-band}, which further favors the integer regime.

\smallskip
{\it Additional intervalley modes in rhombohedral graphene}---Beyond the intravalley spin mode discussed in the main text, a spin-valley-polarized state in rhombohedral multilayer graphene supports two additional intervalley collective modes: one involving a valley flip and the other involving both spin and valley flips. Their gaps are schematically given by $\lambda_{\rm SOC}+ \Delta _{\rm v}$ and  $g\mu_B B + \Delta_{\rm sv}$, respectively, where $\Delta _{\rm v}$ and $\Delta _{\rm sv}$ denote interaction contributions. For the mixed spin-valley mode, measurements in rhombohedral tetralayer graphene found a large intervalley spin-exchange scale of $ 6.5\, {\rm meV} \sim 75 \, {\rm K}$ in a neighboring half-metallic phase \cite{auerbach2025isospin}. Although this scale cannot be directly identified with $\Delta_{\rm sv}$, its magnitude suggests that this mode is unlikely to contribute appreciably at experimentally relevant temperatures $T_c \sim 100$--$340\, {\rm mK}$.  We further assume that the valley-flip mode is also gapped on a scale much larger than $k_BT_c.$

\section{Validity of the low-temperature free-energy expansion}
\label{app:validity}

In deriving the low-$T$ free energies and transition temperatures $T_c$ discussed in cases (i)--(v), two assumptions were implicitly made: (1) only the leading thermal contributions are retained, while subleading power-law and activated contributions are neglected at $T_c$; (2) where gapless modes are present, the asymptotic low-energy Goldstone dispersions of the $T=0$ ordered states are used to evaluate the finite-$T$ free energies. We discuss the validity of these two assumptions below.

\smallskip
(1) {\it Hierarchy of thermal contributions}---The truncation of the low-$T$ free-energy expansion is controlled 
provided that $T_c$ lies sufficiently below the scales at which the neglected contributions  become important. For case (i), this requires 
\begin{align}
\label{eq:self-consistency_Coulomb}
       \frac{(k_B T_c)^{2}}{\hbar J } 
       \gg 
       \frac{(k_B T_c)^{7/3}}{(\hbar A_{\frac 3 2})^{4/3}} 
       , 
       \frac{(k_B T_c)^{3}}       
       {(\hbar v)^{2}},
       \ldots,
\end{align}
together with  $k_BT_c \ll \Delta^{\rm th}_I, \Delta^{\rm th}_F$ so that activated contributions are negligible.  For case (ii), the net $O(T^2)$ contribution must also dominate all the subleading corrections, such as $O(T^3)$ antiferromagnetic-magnon contribution,
\begin{align}
    \frac{(k_BT_c)^2}{\hbar}
    \left|
    \frac{1}{A_2}-\frac{1}{J}
    \right|
    \gg
    \frac{(k_BT_c)^3}{(\hbar v)^2},
\end{align}
again with  $k_BT_c \ll \Delta_I^{\rm th},\Delta_F^{\rm th}$. Near $A_2=J$, the leading $T^2$ terms nearly cancel, so subleading contribution can become important and may need to be retained. For case (iii), we additionally require $k_BT_c \ll \hbar \omega_{\rm pin}$ so that the pinned phonon contribution is thermally suppressed. Analogous self-consistency conditions apply to  cases (iv)--(v).

\smallskip
(2) {\it Validity of the asymptotic Goldstone dispersions}---The free-energy expressions 
(\ref{eq:spin-wave-free-energy-integer_spin-degenerate}, 
\ref{eq:spin-wave-free-energy-fractional_spin-degenerate}, 
\ref{eq:magnetophonon-free-energy}) 
in the main text were obtained using the asymptotic Goldstone dispersions of the $T=0$ ordered states. At finite temperature, these expressions remain applicable provided that the momenta of the thermally excited modes lie above the finite-$T$ infrared (IR) cutoffs set by the inverse correlation lengths and below the ultraviolet (UV) scales at which the dispersions deviate from their asymptotic low-energy forms.

For the spin sector in cases (i)--(iii), although $SU(2)$ symmetry cannot be spontaneously broken at any $T>0$ in 2D \cite{mermin1966absence}, the  finite-$T$ spin-correlation lengths of the spin-textured eIQH and ferromagnetic FQH regimes are exponentially large at low temperature, $\xi_{a}^{\rm sp}(T) \sim  \xi_{a}^{(0)} e^{\rho_{a }/k_BT }$, where $\rho_{a}$ is the appropriate (energy-valued) spin stiffness  for  $a=I,F$; for example, $\rho_{F} \sim  \hbar J /\ell_B^2 $   \cite{chubukov1994universal,read1995continuum}. The $T=0 $ dispersions therefore control the thermodynamics provided the thermally excited modes have momenta much larger than the corresponding inverse correlation lengths $[\xi_{I,F}^{\rm sp}]^{-1}$. 
For the $U(1)$-breaking spin-textured eIQH states in cases (iv) and (v-a), the corresponding spin-wave description applies in the quasi-long-range-ordered regime below $T_{\rm BKT}$; above $T_{\rm BKT}$, the finite spin-correlation length provides the analogous IR cutoff.
For the crystalline sector, below the melting temperature $T< T_{\rm melt}$, the integer crystal has quasi-long-range translational order with well-defined gapless elastic modes~\cite{halperin1978theory,nelson1979dislocation,young1979melting}. Above $T_{\rm melt}$, the translational order becomes short-ranged with a finite correlation length $\xi_I^{\rm cr}$; the long-wavelength elastic description then applies only for momenta $k_I^{\rm ph} \gg 1/\xi_I^{\rm cr}$.

In addition, all relevant thermal momenta must remain below the UV scales at which the corresponding dispersions deviate from the asymptotic low-energy forms $\omega \sim k^ p$: we denote these scales by $k_{D,I }^{\rm sw}$,  $k_{D,I }^{\rm ph}$ and $k_{D,F}^{\rm sw}$. Defining the  characteristic thermal momentum by $\hbar\omega(k_T) \sim  k_BT$, we obtain  
\begin{align}
\label{eq:thermal-momenta}
    &k_{I}^{\rm sw}(T)
    \sim 
    \frac{k_BT}{\hbar v},
    \ \ \ \ \ 
    k_{F}^{\rm sw}(T)
    \sim 
    \sqrt{\frac{k_BT}{\hbar J}},
    \ \ \ \ \ 
    k^{\rm ph}_I(T)
    \sim
    \left(
    \frac{k_BT}{\hbar A_{\frac 3 2}}
    \right)^{\frac 2 3}\!.    
\end{align}
For case (i), the validity of the leading transition temperature \eqref{eq:Tc_Coulomb_spin-degenerate} additionally requires 
\begin{align}
    \label{eq-app:IR-UV-condition}
    1/ \xi_I^{\rm sp } (T_c)
    \ll  
    k_{I}^{\rm sw}(T_c)
     \ll 
    k_{D,I}^{\rm sw},
     \ \ \ \ \ 
    1/ \xi_F^{\rm sp}(T_c) 
    \ll 
    k_{F}^{\rm sw}(T_c)
     \ll 
    k_{D,F}^{\rm sw},
    \ \ \ \ \ 
    1/\xi_I^{\rm cr} (T_c) 
    \ll
    k^{\rm ph}_I(T_c)
    \ll 
    k_{D,I }^{\rm ph },
\end{align}
where it is understood that $[\xi_I^{\rm cr} (T)]^{-1 } = 0$ for $T<T_{\rm melt}$. 
For case (ii), the same conditions apply, except that
$k_I^{\rm ph}
\sim
\left [
{k_BT}/{\hbar A_2}
\right ]^{1/2}.$  If the short-range interaction arises from screening by a metallic gate at distance $d$, $V(\v q) \propto (1-e^{-2qd})/q $ becomes  effectively short-ranged for $qd \ll 1$. Therefore, the quadratic-magnetophonon treatment additionally requires $k_I^{\rm ph}(T_c) d  \ll 1$; otherwise, there is a crossover toward the Coulomb regime of case (i).  For case (iii), only the first two spin-sector conditions of~\eqref{eq-app:IR-UV-condition} need be satisfied. For case (iv), the corresponding magnetophonon condition [last condition of \eqref{eq-app:IR-UV-condition}] applies, together with the spin-wave condition when the eIQH  state is spin-textured. For case (v-a), only the spin-textured eIQH spin-wave condition [first condition of~\eqref{eq-app:IR-UV-condition}] remains when the phonon is strongly pinned, while case (v-b) has no gapless mode under the assumptions made above.

\appendix
\onecolumngrid

\section{Free-Energy Density of $\omega(\v k) = A_p k^p$ Mode}
\label{app:free-energy}
Here, we calculate the free-energy density  of the gapless mode with dispersion relation $\omega(\v k) = A_p k^p$. $p=2 $ is relevant to the  magnetophonon mode of the Wigner crystal (WC) in the short-range interacting case or to the ferromagnetic spin waves. $p=\frac 3 2$ is relevant to the magnetophonon mode in the presence of long-range Coulomb interactions. $p=1 $ is relevant to linear phonon branches of the WC  or to the antiferromagnetic magnons. 

First consider a single bosonic mode with energy $\hbar \omega$. Its partition function is given by 
\begin{align}
    Z_\omega  = \sum_{n=0}^\infty e^{-\beta \hbar \omega  (n+\frac 1 2 )}  =  \frac{e^{-\beta \hbar \omega/2}}{1 - e^{-\beta \hbar \omega}}.
\end{align}
The free energy of a  single bosonic mode with energy  $\hbar \omega $ is   
\begin{align}
     f_\omega  = -k_B T \ln Z_\omega = \frac {\hbar \omega}{2 } + k_B T \ln (1 - e^{-\beta \hbar \omega }).
\end{align}
We measure the free energy relative to its zero-temperature value and therefore drop the first term. To obtain the full free energy density, we sum over all the modes
\begin{align}
    f^{(p)} 
    &=
    k_B T 
    \int 
    \frac {d^2 \v k}{(2\pi)^2 } 
    \ln 
    \left (
    1 - e^{-\beta \hbar A_p k^p }  \right ) 
    = 
    \frac {k_B T}{2\pi }
    \int kdk \,  
    \ln \left (
    1 - e^{-\beta \hbar A_p k^p }  
    \right ) 
    \nonumber \\
    &= 
    \frac {k_B T}{2\pi p}
    (\beta \hbar A_p)^{-\frac 2 p}
    \int_0 ^\infty dx\,  
    x^{\frac 2 p -1 } 
    \ln 
    \left (
    1- e^{-x} 
    \right )
    = 
    - C_p 
    \frac
    {(k_BT)^{1+\frac 2 p}}
    {(\hbar A_p)^\frac 2 p}
    ,
     \\
    C_p 
    &=
    \frac 
    {\Gamma(1+\frac 2 p) \zeta(1+ \frac 2 p)}
    {4\pi }
\end{align}
where in the second line we changed variables to $x= (\beta \hbar A_p)k^p $. $\Gamma$ and $\zeta$ are the gamma and Riemann zeta functions, respectively.  Here, the upper momentum cutoff in the integral was sent to $\infty$; this is a safe approximation as long as the temperature is much smaller than the maximum energy of the relevant mode. For $p=1, \frac 3 2, 2$, we thus have
\begin{align}
    f^{(p)} = 
    \begin{cases}
            -C_1
            \frac
           {(k_BT)^{3}}
           {(\hbar A_{1})^2 }  
           & \text{for } p= 1
            \\
           - C_{3/2}
           \frac
           {(k_BT)^{7 / 3}}
           {(\hbar A_{3/ 2})^{4 /3} }  
           & \text{for } p= \frac 3 2 
           \\
          - C_2
           \frac{(k_B T)^{2}}{\hbar A_{2} } 
           & \text{for } p= 2
    \end{cases},
\end{align}
where $C_1 = 0.191313$, $C_{3/2} = 0.134083 $ and $C_ 2 = \frac \pi {24} $.

\section{Phonons of a Wigner Crystal}
\label{app:magneto-phonon-plasmon}
Here, we review the phonon spectrum of a  WC in the presence of a magnetic field $B$. At $B=0$, there are two phonon modes, longitudinal and transverse. For short-range interactions, both are linear at small $k$; for long-range Coulomb interactions, the longitudinal mode is plasmonic, $\omega_L \sim \sqrt{k}$,  while the transverse mode remains linear. A magnetic field mixes the longitudinal and transverse modes, resulting in a gapless magnetophonon and a gapped magnetoplasmon.

Let us first express the magnetophonon and magnetoplasmon modes at $\v B = B \hat z $ in terms of the zero-field longitudinal $\omega_L(\v k)$ and transverse $\omega_T(\v k)$ phonon modes. Let $\omega_c =   {eB}/{m}$ be the cyclotron frequency and 
\begin{align}
    \v u(\v k,t) = u_L \hat k + u_T (\hat z \times \hat k)  = \frac 1 {\sqrt{L^2}}\sum_{i} e^{i\v k \cdot \v R_i} \v u_i( t)
\end{align}
be the Fourier transform of the displacement field $\v u_i( t)$ at site $\v R_i$ of the WC [$\v R_i$ labels $L^2$ WC sites]. Then, the equation of motion at $B>0 $ becomes
\begin{align}
    m\ddot {\v u}  = -m\omega_L^2 u_L  \, \hat k -m\omega_T^2 u_T \,  (\hat z \times \hat k ) -e \dot {\v u} \times \v B,
\end{align}
where the last term is the Lorentz force term. Taking $\v u \sim e^{-i\omega(\v k) t}$, we obtain
\begin{align}
    \begin{pmatrix}
        \omega_L^2 - \omega^2  & -i\omega \omega_c \\
         i\omega \omega_c & \omega_T^2 - \omega^2 
    \end{pmatrix}
    \begin{pmatrix}
        u_L \\ u_T
    \end{pmatrix}
    = 0.
\end{align}
This gives $(\omega_L^2 - \omega^2 )(\omega_T^2 - \omega^2 ) -\omega^2 \omega_c ^2 =0$, and therefore,
\begin{align}
\label{eq-app:magneto-phonon-plasmon-exact}
    \omega^2_\pm (\v k) 
    = 
    \frac 1 2 
    \left [ 
    \omega_L^2 + \omega_T^2 + \omega_c^2
    \pm
    \sqrt {
(\omega_L^2 + \omega_T^2 + \omega_c^2)^2 
-4 \omega_L^2  \omega_T^2 
    }
    \right ],
\end{align}
where the lower branch $\omega_-$ is called  the magnetophonon and the upper branch $\omega_+ $ the magnetoplasmon. In the long-wavelength limit, $\omega_c \gg \omega_L(\v k),\omega_T(\v k)$, and \eqref{eq-app:magneto-phonon-plasmon-exact} becomes
\begin{align}
\label{eq-app:magneto-phonon-plasmon-small-k}
    \omega_-(\v k) 
    \approx 
    \frac{\omega_L(\v k)\omega_T(\v k)}{\omega_c},
    \ \ \ \ \ 
    \omega_+(\v k) 
    \approx 
    \omega_c + 
    \frac{\omega_L^2 (\v k)+ \omega_T^2 (\v k)}{2\omega_c}.
\end{align}

\subsection{Short-Range Interactions}
When the WC arises from short-range interactions, longitudinal and transverse phonons are both linearly dispersing at long wavelength:
\begin{align}
    \omega_L(\v k) \approx  c_L k,
    \ \ \ \ \ 
    \omega_T(\v k) \approx  c_T k,
\end{align}
where  $c_L$ and $c_T$ are the longitudinal and transverse sound velocities, respectively, which depend on the details of the interaction. They are related to the bulk and shear moduli $K$ and $\mu$ as \cite{chaikin1995principles}
\begin{align}
    c_L=
    \sqrt{\frac{K+\mu }{m n}}, 
    \ \ \ \ \
     c_T=
     \sqrt{\frac{\mu }{m n}}, 
\end{align}
where $n $ is the number of WC electrons per unit area. Therefore,
\begin{align}
\label{eq-app:SR-magneto-phonon-plasmon}
    \omega_-(\v k) 
    \approx 
    A_{2}
    k ^2, 
    \ \ \ \ \ 
    \omega_+(\v k)
    \approx 
    \omega_c 
    + \frac{c_L^2 + c_T^2}{2\omega_c} k ^2,
\end{align}
\begin{align}
      A_{2}= \frac{{c_L c_T}}{\omega_c } = \frac{\sqrt{\mu(K+ \mu ) }}{eB n}.
\end{align}

\subsection{Long-Range Coulomb Interaction}
In the presence of the Coulomb interaction, $V(\v r) = \frac {e^2 }{4\pi \epsilon  | \v r|} $, the longitudinal phonon becomes plasmon-like \cite{bonsall1977some} whereas the transverse phonon is linearly dispersing 
\begin{align}
    \omega_L(\v k) 
    \approx 
    \left (\frac{ n  e^2 }{2\epsilon  m} k \right )^{1/2},
    \ \ \ \ \ 
    \omega_T(\v k) \approx c_T k,
\end{align}
where the transverse phonon velocity is again related to the shear modulus as $c_T = (\frac \mu {m n })^{1/2}$.  In the presence of a magnetic field, the magnetophonon and magnetoplasmon dispersions \eqref{eq-app:magneto-phonon-plasmon-small-k} thus become
\begin{align}
\label{eq-app:Coulomb-magneto-phonon-plasmon}
    \omega_- (\v k) \approx  A_{\frac 3 2} k^{\frac 3 2},
    \ \ \ \ \
    \omega_+  (\v k) 
    \approx 
    \omega_c +  
    \frac{1}{2\omega_c}
    \frac{ n e^2 }{2\epsilon m} k ,
\end{align}
\begin{align}
    A_{\frac 3 2} = 
    \frac {c_T} {\omega_c}
\sqrt {
\frac{  n e^2 }{2 \epsilon m}
}     = \frac 1 {eB} \sqrt {\frac {e^2 }{2\epsilon } \mu }.
\end{align}
For a pure Coulomb interaction, the shear modulus is given by \cite{bonsall1977some}
\begin{align}
    \mu \approx 0.245 \frac{e^2 }{4\pi \epsilon } n ^{\frac 3 2},
\end{align}
and the stiffness coefficient scales as $A_{\frac 3 2 } \propto n^{\frac 3 4}$.

\section{Magnetoroton Free Energy}
\label{app:magnetoroton}
Here, we calculate the free-energy density of a gapped magnetoroton mode, whose minima are at finite $k$. We consider (a) a ring of minima, appropriate to a rotationally invariant LL or to a Chern-band case with weak warping effects, and (b)  a finite set of symmetry-related minima produced by appreciable warping.

(a) When the minima occur around a ring of $\v k$ points, the magnetoroton dispersion can be expanded as 
\begin{align}
    \epsilon_R (k ) 
    =
    \Delta _R  
    +
    \frac 
    {\hbar^2 (k-k_R)^2}
    {2m_R } 
    +\ldots. 
\end{align}
The free-energy density is 
\begin{align}
\label{eq-app:magnetoroton-free-energy-continuum}
    f_{\rm R}
    &= 
    k_B T 
    \int 
    \frac {d^2 \v k}{(2\pi)^2 }
    \ln 
    \left ( 
    1 - 
    e^{-\epsilon_R(k) / k_BT }
    \right )  
    \approx  
    k_B T 
    \int 
    \frac {d^2 \v k}{(2\pi)^2 }
    \ln 
    \left ( 
    1 - 
    e^{-\Delta_R / k_BT }
    e^{
    -\frac
    {\hbar^2 (k-k_R)^2 }
    {2m_R k_BT }
    }
    \right ) 
    \nonumber \\
    &\approx  
     \frac{k_R\,  k_BT }{2\pi }
    \int_{-\infty}^\infty d q
    \ln
    \left [
    1 - 
    \exp 
    (-\Delta_R / k_BT  
    - 
    \frac{\lambda_T^ 2  q^2}{4\pi } )
    \right ]
    = 
    - \frac{k_R\,  k_BT }{2\pi }
    \int_{-\infty}^\infty d q
    \sum_{n= 1 }^\infty 
    \frac{e^{-n\Delta_R /k_B T}}{n}
    e^{-\frac{ n \lambda_T ^2 q^2}{4\pi} }
    \nonumber \\ &
    =
    - \frac{k_R\,  k_BT }{\lambda_T }
    \sum_{n= 1 }^\infty 
    \frac
    {e^{-n\Delta_R /k_B T}}
    {n^{3/2 }}
    =
    - \gamma_R (k_BT)^{3/2}
    {\rm Li}_{3/2 } 
    (e^{-\Delta_R /k_B T}),
\end{align}
where $\lambda_T = h /\sqrt{2\pi m_R  k_B T }$ is the de Broglie thermal length, $\gamma_R \equiv \frac{k_R \sqrt {m_R}}{\sqrt {2\pi }\hbar }$, and  ${\rm Li}_{s}(z) \equiv \sum_{n=1}^\infty  z^n / n^s $ is the polylogarithm of order $s.$ In the second line, $kdk $ is replaced by $(k_R + q)dq \approx k_R dq $, assuming  that the thermally excited states are concentrated close to $k = k _R $. This narrow-shell approximation requires $k_BT \ll E_R \equiv \hbar^2 k_R^2 /2m_R $.
When the temperature is much lower than the magnetoroton gap, $k_BT \ll \Delta_R $, the free-energy density has the activated form, 
$f_R \sim 
- \gamma_R (k_BT)^{3/2
}
e^{-\Delta_R /k_B T}.$ This is parametrically larger than the case (b) where minima occur at a finite number of points, whose low-temperature scaling is $O(T^2 \exp[-\Delta _R/k_BT])$. This reflects  the diverging density of states associated with the ring of minima: $\rho_R (E ) \propto 1/\sqrt{E- \Delta _R } $.

(b) When the continuous rotation symmetry is no longer a good approximation, the magnetoroton dispersion around the ring $k = k_R $ is strongly warped. For example, if the system has $C_n$ symmetry, an additional $n$-gonal warping term must be added ($u_n > 0$):
\begin{align}
\label{eq-app:C3_magnetoroton}
    \epsilon_R (k,\theta) 
    =
    \Delta _R  
    +
    \frac 
    {\hbar^2 (k-k_R)^2}
    {2m_R } 
    + u_n
    \left \{  1- 
    \cos[n(\theta-\theta_n) ]
    \right \} 
    +\ldots. 
\end{align}
The approximate description in (a) is appropriate when the warping energy
\begin{align}
\label{eq-app:warping-energy}
    W
    = 
    \max_\theta 
    \epsilon_R(k_R , \theta)
    -
    \min_\theta 
    \epsilon_R(k_R , \theta) \approx 2 u_n 
\end{align}
satisfies $W \ll k_B T$. 
On the other hand, when $W  \gg k_B T$, the dispersion can be expanded around the $N_R$ inequivalent symmetry-related minima:
\begin{align}
    \epsilon_{R,j } (k,\theta) 
    \approx 
    \Delta _R  
    +
    \frac 
    {\hbar^2 (k -k_R) ^2}
    {2m_R} 
    +
    \frac 
    {\hbar^2 k_R ^2}
    {2m_\theta} 
    \left (
    \theta 
    - 
    \frac {2\pi j }{n} 
    - 
    \theta_ n
    \right )^2 
    , 
    \ \ \
    j=0,\ldots, N_R-1 
\end{align}
where  $\v k = k  (\cos \theta, \sin \theta )$, and  $m_\theta$ is related to $u_n$ by $n^2 u_n = \hbar^2 k_R^2 /m_\theta $. Note that $N_R$ can be smaller than $n$ when the minima $k_R (\cos (\frac{2\pi j }{n} + \theta_n), \sin (\frac{2\pi j }{n} + \theta_n))$ for $j=0,\ldots, n-1$ are related to one another by reciprocal lattice vectors; otherwise, $N_R = n$. Repeating the calculation above, the free-energy density becomes
\begin{align}
\label{eq-app:magnetoroton-free-energy-discrete}
    f_R ^{\rm disc}
    =
    - N_R
    \frac
    {\sqrt{m_R m_\theta}}
    {2\pi \hbar^2 }
    (k_BT)^2 
    {\rm Li}_2 
    (e^{-\Delta_R /k_BT}),
\end{align}
which is $O(T^2 \exp[-\Delta _R/k_BT])$ at low temperatures.

\fi

%TC:endignore
\twocolumngrid

\bibliography{main.bib}

\end{document}